\documentclass[a4paper,11pt]{article}
\usepackage{jinstpub} 
\usepackage{enumitem}

\title{\boldmath New method for clustering thresholds determination in Microstrip Silicon Detector}

\author[aj,c,*]{S. Mazzolani,\note[*]{Corresponding author.}}
\author[a]{I. Mattei,}\author[b]{A. Alexandrov,}\author[c]{B. Alpat}
\author[c]{G. Ambrosi,}
\author[d,e]{S. Argirò,}
\author[c]{M. Barbanera,}
\author[e]{N. Bartosik,}
\author[a]{G. Battistoni,}
\author[f,g]{M.G. Bisogni,}
\author[h,b]{V. Boccia,}
\author[n]{G. Butella,}
\author[c]{M. Caprai,}
\author[e]{F. Cavanna,}
\author[e]{P. Cerello,}
\author[f,g]{E. Ciarrocchi,}
\author[am]{N. D'Ambrosio,}
\author[h,b]{G. De Lellis,}
\author[h,b]{A. Di Crescenzo,}
\author[m,l]{M. Dondi,}
\author[n]{M. Donetti,}
\author[a]{Y. Dong,}
\author[h,o,ak]{M. Durante,}
\author[i,j]{R. Faccini,}
\author[e]{V. Ferrero,}
\author[p]{C. Finck,}
\author[e]{E. Fiorina,}
\author[b]{M. Francesconi,}
\author[m,l]{M. Franchini,}
\author[q,j]{G. Franciosini,}
\author[r,s]{G. Galati,}
\author[g]{L. Galli,}
\author[c]{M. Ionica,}
\author[b,h]{A. Iuliano,}
\author[t,c]{K. Kanxheri,}
\author[e]{B. Kharpuse,}
\author[g]{A.C. Kraan,}
\author[h,b]{A. Lauria,}
\author[w,e]{E. Lopez Torres,}
\author[q,j]{M. Magi,}
\author[l,m]{A. Manna,}
\author[x,j]{M. Marafini,}
\author[am]{S. Masci,}
\author[g]{M. Massa,}
\author[l,m]{C. Massimi,}
\author[l]{A. Mengarelli,}
\author[n]{A. Mereghetti,}
\author[q,j]{R. Mirabelli,}
\author[g]{A. Moggi,}
\author[aa,b]{M.C. Montesi,}
\author[y,z]{M.C. Morone,}
\author[f,g]{M. Morrocchi,}
\author[a]{S. Muraro,}
\author[e]{N. Pastrone,}
\author[q,j]{V. Patera,}
\author[e]{F. Pennazio,}
\author[l,m]{C. Pisanti,}
\author[c,ab]{P. Placidi,}
\author[n]{M. Pullia,}
\author[i,an,j]{F. Quattrini,}
\author[ac,e]{L. Ramello,}
\author[o]{C. Reidel,}
\author[l,m]{R. Ridolfi,}
\author[ad]{L. Sabbatini,}
\author[c,t]{L. Salvi,}
\author[ad]{C. Sanelli,}
\author[q,j]{A. Sarti,}
\author[ae]{O. Sato,}
\author[n]{S. Savazzi,}
\author[af]{L. Scavarda,}
\author[q,j]{A. Schiavi,}
\author[o]{C. Schuy,}
\author[v]{E. Scifoni,}
\author[c]{G. Silvestre,}
\author[ac,e]{M. Sitta,}
\author[d,e]{B. Spadavecchia,}
\author[l]{R. Spighi,}
\author[ad]{E. Spiriti,}
\author[i,j]{L. Testa,}
\author[b]{V. Tioukov,}
\author[ad]{S. Tomassini,}
\author[v,ai]{F. Tommasino,}
\author[q,j]{M. Toppi,}
\author[j]{G. Traini,}
\author[ad]{A. Trigilio,}
\author[m,l]{G. Ubaldi,}
\author[l,m]{S. Valentinetti,}
\author[p]{M. Vanstalle,}
\author[m,l]{M. Villa,}
\author[o,al]{U. Weber,}
\author[l,m]{R. Zarrella,}
\author[l,m]{A. Zoccoli,}
\author[c]{L. Servoli.}

\affiliation[a]{INFN Section of Milano, Milano, Italy}
\affiliation[b]{INFN Section of Napoli, Napoli, Italy}
\affiliation[c]{INFN Section of Perugia, Perugia, Italy}
\affiliation[d]{University of Torino, Department of Physics, Torino, Italy}
\affiliation[e]{INFN Section of Torino, Torino, Italy}
\affiliation[f]{University of Pisa, Department of Physics, Pisa, Italy}
\affiliation[g]{INFN Section of Pisa, Pisa, Italy}
\affiliation[h]{University of Napoli, Department of Physics 'E. Pancini', Napoli, Italy}
\affiliation[i]{University of Rome La Sapienza, Department of Physics, Rome, Italy}
\affiliation[j]{INFN Section of Roma 1, Rome, Italy}
\affiliation[k]{University of Foggia, Foggia, Italy}
\affiliation[l]{INFN Section of Bologna, Bologna, Italy}
\affiliation[m]{University of Bologna, Department of Physics and Astronomy, Bologna, Italy}
\affiliation[n]{CNAO National Center for Oncological Hadrontherapy, Pavia, Italy}
\affiliation[o]{Biophysics Department, GSI Helmholtzzentrum für Schwerionenforschung, Darmstadt, Germany}
\affiliation[p]{Université de Strasbourg, CNRS, IPHC UMR 7871, F-67000 Strasbourg, France}
\affiliation[q]{University of Rome La Sapienza, Department of SBAI, Rome, Italy}
\affiliation[r]{University of Bari, Department of Physics, Bari Italy}
\affiliation[s]{INFN Section of Bari, Bari, Italy}
\affiliation[t]{University of Perugia, Department of Physics and Geology, Perugia, Italy}
\affiliation[u]{University of Miami, Radiation Oncology, Miami, FL, United States}
\affiliation[v]{Trento Institute for Fundamental Physics and Applications, INFN, Trento, Italy}
\affiliation[w]{CEADEN, Centro de Aplicaciones Tecnologicas y Desarrollo Nuclear, Havana, Cuba}
\affiliation[x]{Museo Storico della Fisica e Centro Studi e Ricerche Enrico Fermi, Rome, Italy}
\affiliation[y]{University of Rome Tor Vergata, Department of Physics, Rome, Italy}
\affiliation[z]{INFN Section of Roma Tor Vergata, Rome, Italy}
\affiliation[aa]{University of Napoli, Department of Chemistry, Napoli, Italy}
\affiliation[ab]{University of Perugia, Department of Engineering, Perugia, Italy}
\affiliation[ac]{University of Piemonte Orientale, Department for Sustainable Development and Ecological Transition, Vercelli, Italy}
\affiliation[ad]{INFN Laboratori Nazionali di Frascati, Frascati, Italy}
\affiliation[ae]{Nagoya University, Department of Physics, Nagoya, Japan}
\affiliation[af]{ALTEC, Aerospace Logistic Technology Engineering Company, Torino, Italy}
\affiliation[ai]{University of Trento, Department of Physics, Trento, Italy}
\affiliation[aj]{University of Camerino, Department of Physics, Camerino, Italy}
\affiliation[ak]{Institute of Condensed Matter Physics, Technische Universität Darmstadt, Darmstadt, Germany}
\affiliation[al]{Institute of Medical Physics and Radiation Protection (IMPS), University of Applied Sciences, Giessen, Germany}
\affiliation[am]{INFN, Gran Sasso National Laboratories, Assergi (L’Aquila), Italy}
\affiliation[an]{Specialty School of Medical Physics, Sapienza, Rome, Italy}

\collaboration{The FOOT Collaboration}
\emailAdd{sofia.mazzolani@unicam.it}

\abstract{The Microstrip Silicon Detector (MSD) is one of the subsystems of the FragmentatiOn Of Target (FOOT) experiment whose goal is to measure double differential nuclear fragmentation cross sections for applications in particle therapy and radioprotection in space. The MSD, composed of six 150 $\mu$m-thick silicon sensors arranged in three X-Y measuring planes, is part of the FOOT experiment tracking region. In this work, we propose a new method to set the MSD thresholds for the clustering algorithm independently of the other detectors of the experiment. The obtained values can be used as reference values from which to perform a threshold scan in order to evaluate the single-ion detection efficiency of each MSD sensor and to orient the clustering analysis using tracking information.}

\keywords{Silicon microstrip detectors, Particle tracking detectors, dE/dx detectors, Clustering, Particle therapy}

\begin{document}
\maketitle
\flushbottom

\section{Introduction}
There are numerous experiments in particle and nuclear physics that rely on silicon microstrip detectors to obtain spatial position and energy deposition measurements, and a common challenge across these applications is the development of fast and reliable thresholds setting procedures  for clustering algorithms, which are essential for efficient and rapid signal reconstruction \cite{2002AMS_Tracker,2014CMS_Tracker,FOOT_2021}.
\\The data used in this work have been collected within the FOOT (Fragmentation Of Target) experiment, which is characterized by short data-taking periods during which the time available to evaluate the performance of the various detectors is limited, and by a great variety of datasets in terms of beam configurations and target materials.  In this context, we developed a new approach that allows signal thresholds setting in microstrip detectors without relying on the alignment algorithm of the whole experimental setup nor on the performances of other detectors. The thresholds obtained with this method serve as a starting reference point around which to orient the clustering study based on tracking information.
\subsection{The FOOT experiment} 
\begin{figure}[htbp]
    \centering
    \includegraphics[width=0.7\textwidth]{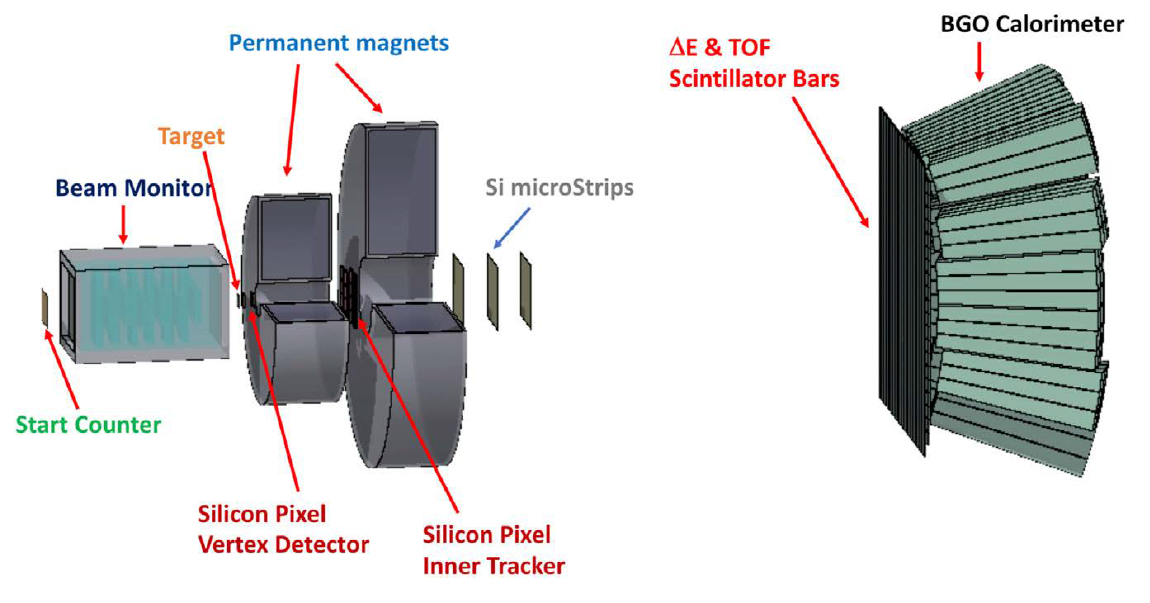}
    \caption{Schematic view of the FOOT Magnetic Spectrometer setup \cite{FOOT_2021}.}
    \label{FOOT}
\end{figure}

The Microstrip Silicon Detector (MSD) is part of the Magnetic Spectrometer of the FragmentatiOn Of Target (FOOT) experiment. The FOOT experiment is an INFN project designed for measuring elemental double differential nuclear fragmentation cross sections with respect to emission angle and kinetic energy of the fragments  \cite{FOOT_2021}. These measures are primarily required for Particle Therapy (PT) applications to improve the accuracy of patient's treatment plans, and for Radioprotection in Space (RPS) for the design of shielding systems for astronauts and electronic devices during long-duration space missions. To this aims, the projectile and target fragmentations are measured from the interaction between p, $^{12}$C, $^{16}$O, $^{4}$He beams and carbon, oxygen and hydrogen-rich targets as they are the principal components of the human body. The kinetic energy range in which FOOT operates is $\sim$100-800 MeV/u (up to 1 GeV/u for RPS). The required double differential cross sections measurement accuracy for projectile fragmentation processes is lower than $5\%$.\\FOOT works with two different setups, the Emulsion Spectrometer, with an angular acceptance around the beam axis bigger than 70° in order to detect fragments with Z $\leq$ 3, and the Magnetic Spectrometer, shown in Figure~\ref{FOOT}, with  10° angular acceptance in order to detect fragments with Z $\geq$ 2; the latter setup is about two meters long and is composed by the Interaction region, the Tracking region and the Time-of-Flight and Calorimetric region.
FOOT has been designed to be a multi-detector transportable experiment capable to fit in different experimental rooms where beams of interest for PT and RPS are available. From 2021 to 2025 FOOT has acquired several data at the GSI research center in Darmstadt (DE), at the Heidelberg Ion Beam Therapy (HIT) Center (DE) and at the National Center for Oncological Hadrontherapy (CNAO) in Pavia (IT).\\\\In this work, we focus on the data taking campaign conducted at CNAO in 2024. In particular  we report results on the MSD signal threshold setting from the measurements using proton beams from 70 MeV to 230 MeV in a no target configuration. The reason is that using high-energy protons without a target results in the minimum energy deposition in the detector: therefore, optimizing the thresholds under these conditions ensures the capability to detect even light fragments (Z = 1) at high energies.
\section{Materials and Methods}
\subsection{The Microstrip Silicon Detector (MSD)}
The MSD is one of the detectors of the Tracking region of the Magnetic Spectrometer, which also comprises two silicon pixel detectors and two permanent magnets (see Figure~\ref{FOOT}). The MSD serves multiple purposes: primarily for particle fragments' tracking together with the Vertex and the ToF Wall detectors in order to measure the fragments momentum, but also for the measurement of the fragments energy loss used for their Z identification that can be matched with the ToF Wall Z identification.
\begin{figure}[htbp]
    \centering
    \includegraphics[width=0.7\textwidth]{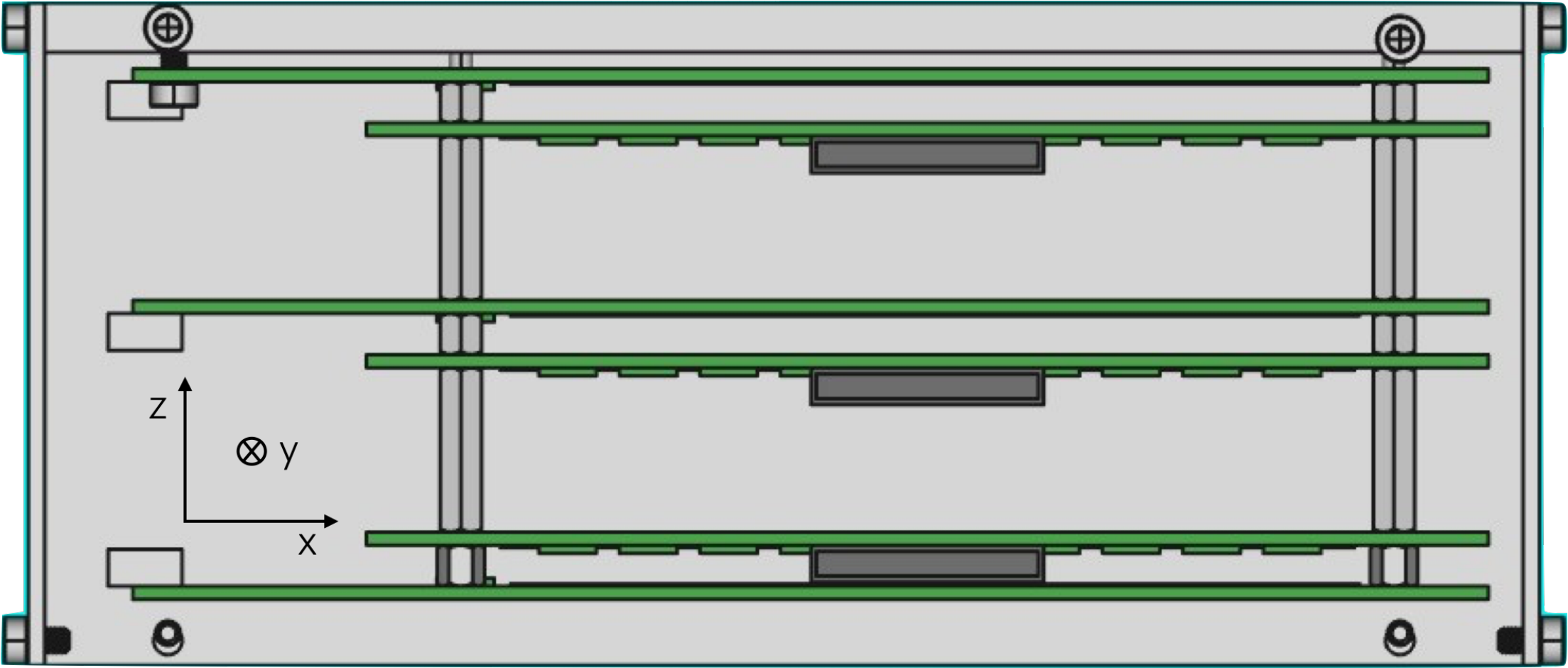}
    \caption{A schematic image of the Microstrip Silicon Detector. It shows a top view of the aluminum box containing the six 150 $\mu$m-thick sensors arranged in pairs, along with the three electronic connectors. The beam direction is along the Z-axis.}
    \label{msd}
\end{figure}
As shown in Figure~\ref{msd}, the MSD consists of three planes separated by a $\sim$ 2 cm gap along the beam direction, each composed by two single-sided silicon sensors arranged orthogonally to provide the X-Y coordinates (the Z-axis is along the beam direction) for a three-dimensional reconstruction of the track. Each silicon sensor has $\sim$ 10 x 10 cm$^2$ active area and  150 $\mu$m thickness, and is equipped with 1920 strips with a width of 50 $\mu$m.  The choice of a reduced sensor thickness with respect to the usual 300 $\mu$m \cite{FOOT-2017} is dictated by the need to reduce nuclear fragmentation outside the target and also to reduce multiple scattering and energy loss along the fragment track. Nevertheless, the signal for high energy protons remains high enough to obtain a  signal-to-noise ratio around 8 \cite{Silvestre_2022}.  For the readout, the choice `1 readout strip - 2 floating strips' is adopted in order to reduce the readout channels while maintaining a spatial resolution of $\sim$ 43 $\mu$m; this results in 640 readout strips per sensor, with a pitch of 150 $\mu$m. The signal is read by 64 channel low-noise/low-power high dynamic range charge sensitive preamplifier-shaper circuits (IDE1140 ASIC), each sensor being equipped with 10 ASIC chips. The analog signals are then transmitted to an ADC board (two ASIC chips for each ADC) to be digitized and then sent to the DE10Nano boards for packing \cite{Kanxheri_2022}. The three MSD planes are enclosed in an aluminum frame (see Figure~\ref{msd}) attached to a mechanical support with six degrees of freedom, allowing for easier mounting and alignment within the complete setup.
\subsection{MSD Calibration}At the acquisition of each event, the readout chain produces a sequence of 640 numerical values for each sensor representing the raw signal, Signal$^{RAW}$, and to properly operate the MSD an accurate preliminary calibration is necessary.\\The first step of the calibration is the calculation and the subtraction of the Pedestal (Ped), that is the average value of the raw signal measured in the absence of particles, for each strip of a sensor. The second step is the calculation and the subtraction of the Common Noise (CN). The CN is a collective signal variation of the single strip analog raw signal readout due to external electromagnetic noise: since the smallest unit of the readout chain is the single ASIC chip, it is calculated for each chip, i.e. every 64 strips, and event by event \cite{Turchetta_1993}.  
The resulting signal for each strip is given by:
\begin{equation}
\text{Signal}_i  = \text{Signal}^{RAW}_i - \text{Ped}_i - \text{CN}_k
\label{eq:raw}
\end{equation}
where \textit{i} is the i-th strip and \textit{k} the k-th chip in a sensor.
The signal distribution for each strip is a gaussian whose sigma is the Single Strip Noise (SSN).
Exploiting the SSN value it is possible to identify noisy and dead strips of each sensor and mask them, which is essential to ensure data integrity, minimize fake hits, stabilize electronics performance and maintain the overall detector efficiency; those bad strips are then excluded from all subsequent operations.
\begin{figure}[htbp]
    \centering
    \includegraphics[width=0.7\textwidth]{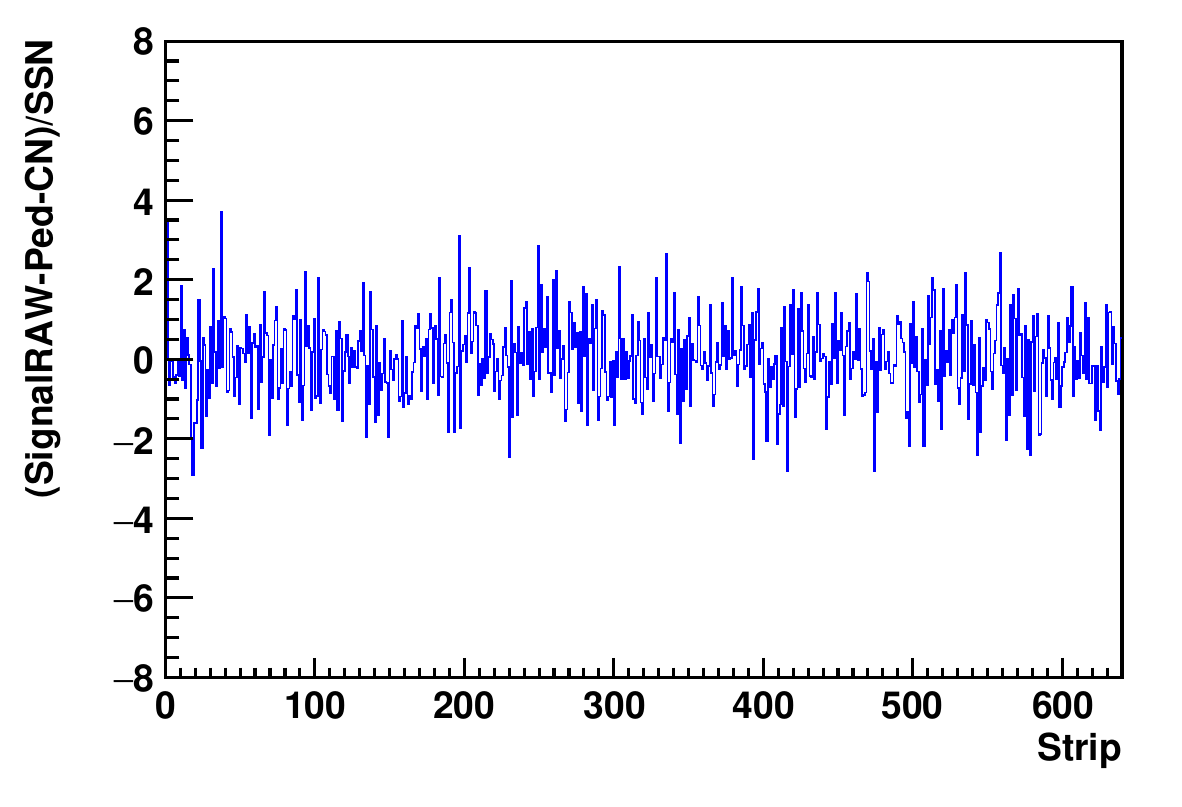}
    \caption{Event display of the quantity $\text{Signal} = \text{Signal}^{RAW} - \text{Ped} - \text{CN}$ normalized to the Single Strip Noise as a function of the strips related to Sensor 4, for a \textit{Calibration Run} of the CNAO2024 data-taking campaign.}
    \label{fig:ev_calib}
\end{figure}Hence, after the calibration, the processed signal for each strip of each sensor oscillates around the zero value as shown in Figure~\ref{fig:ev_calib} where the distribution of one event of $\text{Signal}_i/\text{SSN}_i$ is shown as a function of the readout strip number.
In this case the event display represents one of the six MSD sensors during a data acquisition in absence of particles to which we refer to as \textit{Calibration Run}.
\subsection{Algorithm for setting cluster thresholds}
\label{Algorithm}
\begin{figure}[htbp]
    \centering
    \begin{minipage}[b]{0.45\textwidth}
        \centering
        \includegraphics[width=\textwidth]{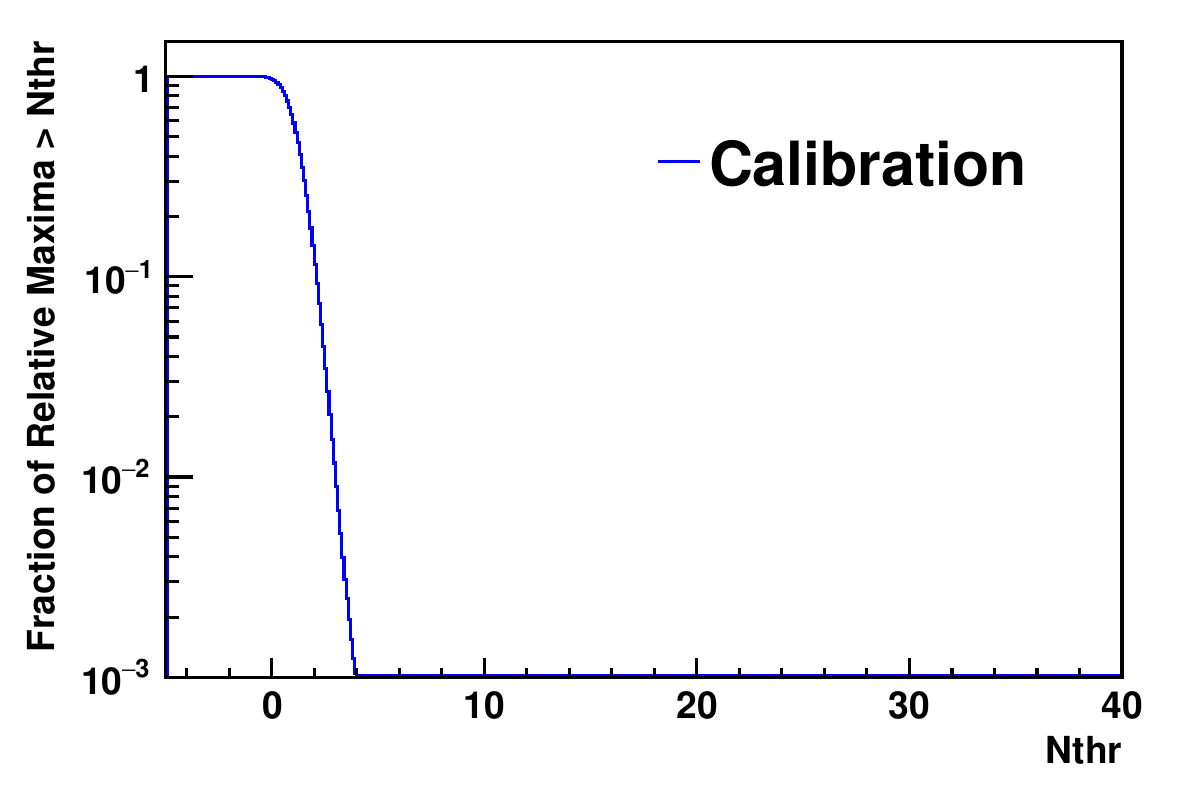}
        \label{fig:subfig_a}
    \end{minipage}
    \hfill
    \begin{minipage}[b]{0.45\textwidth}
        \centering
        \includegraphics[width=\textwidth]{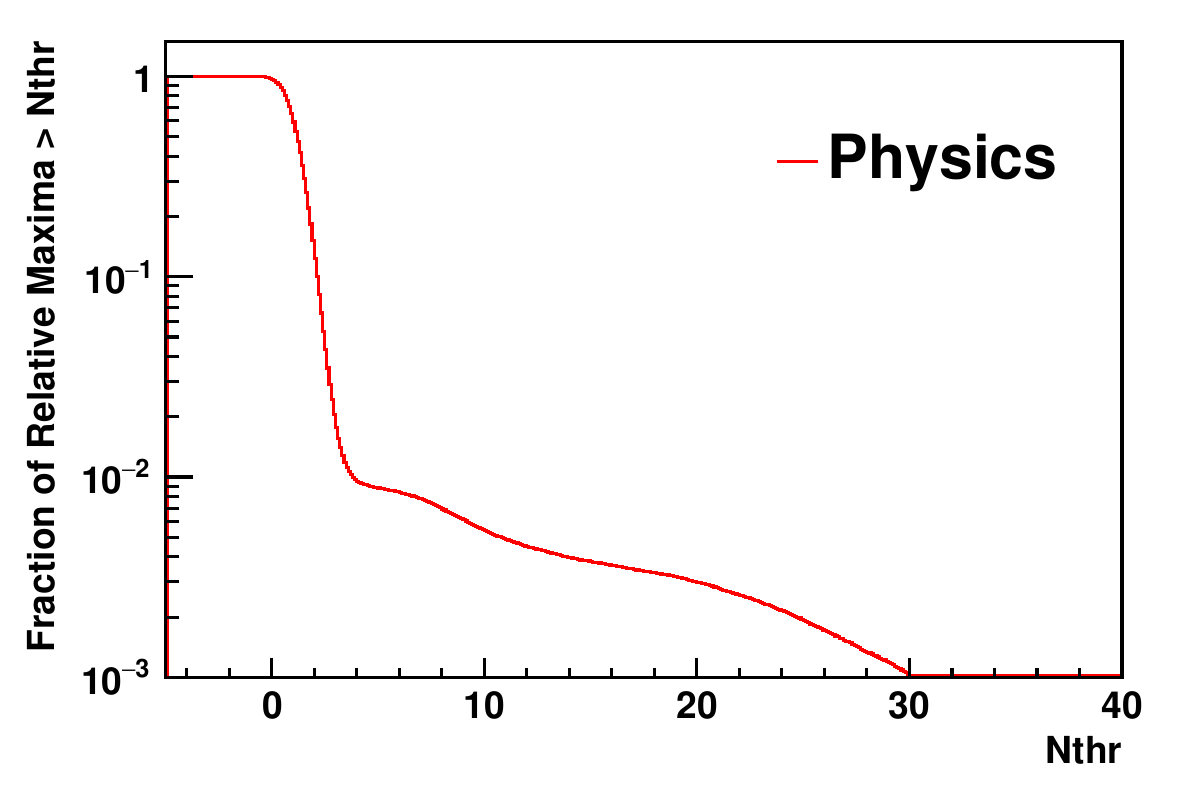}
        \label{fig:subfig_b}
    \end{minipage}
    \caption{Cumulative histograms of Signal$^{Rel_{MAX}}_j/\text{SSN}_j$=$(\text{Signal}^{RAW}_{j} - \text{Ped}_j - \text{CN}_k)/\text{SSN}_j$ when exceeding the threshold variable Nthr, realized using the \textit{strip distance} parameter equal to 1. The curves are normalized to 10$^4$ events, relate to MSD Sensor 4 and have been computed with two runs of CNAO2024 data taking campaign. The Y-axis, in logarithmic scale, represents the fraction of relative maxima above threshold, while the X-axis corresponds to Nthr. \textbf{(Left)} Representation of a \textit{Calibration Run}  (in absence of an incoming particle beam). \textbf{(Right)}  Representation of a \textit{Physics Run} (protons at 230 MeV impinging on the FOOT magnetic spectrometer setup with no target).}
    \label{fig:main_figure}
\end{figure}

When a charged particle passes through the MSD, it is highly unlikely that it crosses just a single strip: more commonly, it crosses the region between two strips, whether they are connected or floating. As a result, the deposited charge is collected in a non-trivial way and the signal is read out often by more than a single strip: therefore a reconstruction among neighbouring strips of the whole signal corresponding to the released energy of a single passing particle is needed in order to determine a cluster \cite{HERAb-2001}. 
To define a cluster two thresholds on the processed signal value of the strips are required: a \textit{Seed Threshold} to identify the presence of a cluster and a \textit{Fired Threshold} to determine its size, that corresponds to the number of strips which contain part of the signal for a given detected charged particle \cite{Silvestre_2022}. 
Setting the threshold values must be done as accurately as possible: too low thresholds may allows undesired noise in the signal leading to false clusters and hence to the degradation of the signal and a more difficult track reconstruction; on the other hand, with too high thresholds there is the risk to miss true clusters and reduce the detection efficiency, hence to lose physical information.\\The adopted strategy to avoid fake clusters involves evaluating the contribution of potential cluster signals collected during \textit{Calibration Runs}. Then a consequent comparison is made with data acquired in the presence of particles, that we refer to as \textit{Physics Runs}.\\Since cluster identification relies on the threshold values themselves, our method actually focuses on evaluating  \textit{Potential Seed Strips}. A \textit{Potential Seed Strip} is the j-th strip whose processed ADC signal (see Equation~\ref{eq:raw}) is a relative maximum among all strip signals belonging to a given sensor. We identify the signal of the j-th \textit{Potential Seed Strip} as Signal$^{Rel_{MAX}}_j$.\\To detect and isolate these \textit{Potential Seed Strips} a dedicated three-points algorithm has been implemented:
\begin{itemize}[noitemsep, topsep=0pt, parsep=0pt, partopsep=0pt]
\item looking for strips with relative maximum signal with respect to the two adjacent ones in a sensor;
\item sorting those candidates in ADC value descending order;
\item starting from the candidate with highest ADC value in the list, filtering out other candidates too close in position to it; closeness is defined, in terms of readout pitch units, by a parameter \textit{strip distance} chosen as a function of the ion under investigation.   
\end{itemize}
For each sensor, once the Signal$^{Rel_{MAX}}_j$ list is obtained on an event-by-event basis, a cumulative histogram is produced by integrating the Signal$^{Rel_{MAX}}_j/\text{SSN}_j$ as a function of a threshold Nthr. 
In particular, as shown in Figure~\ref{fig:main_figure}, the explained algorithm is applied both to a \textit{Calibration Run} (left) and a \textit{Physics Run} (right) of the same session of a data taking campaign (CNAO24) and for the same number of events (10$^4$). The cumulative histograms are then normalized to the total number of events of the sample, and refer to MSD Sensor 4. \cite{Servoli_2010,Servoli_2011}.
The two cumulative distributions have the same behaviour when the threshold is very low.
\section{Data Analysis}
\label{sec:Analysis}
\begin{figure}[htbp]
    \centering
    \includegraphics[width=0.7\textwidth]{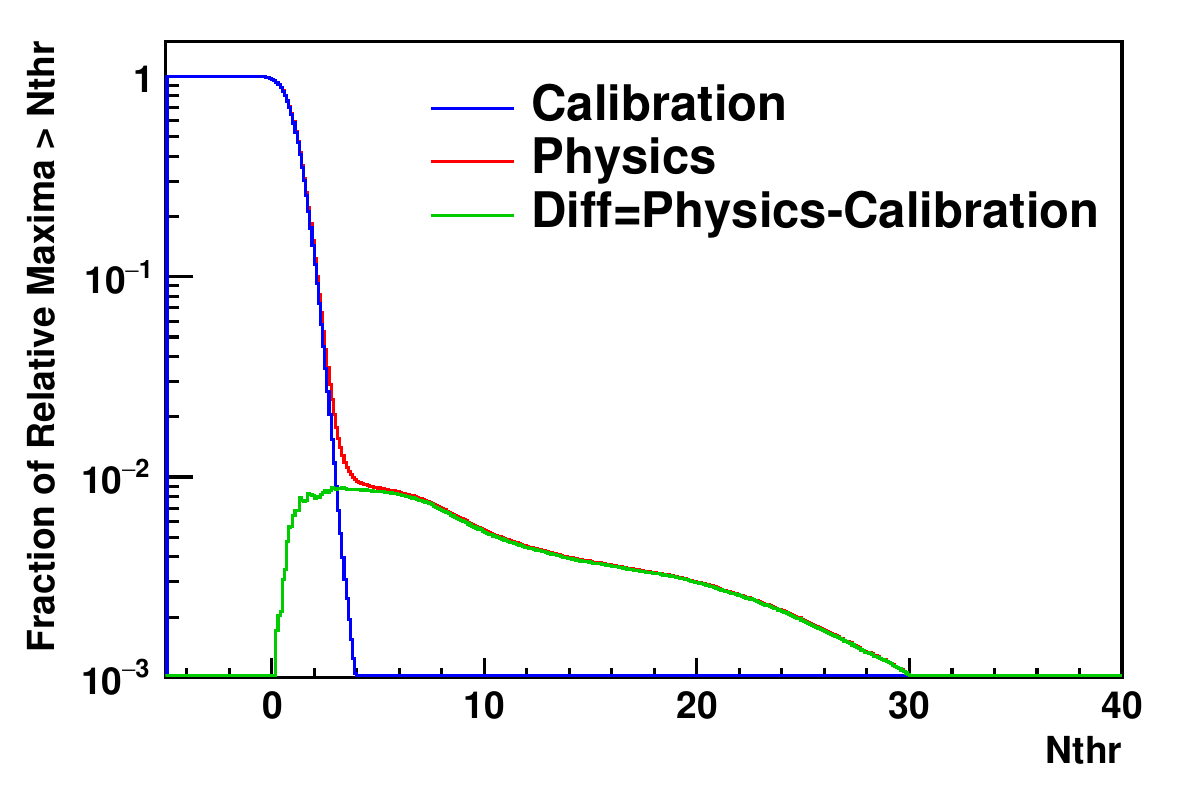}
    \caption{Cumulative histograms of a \textit{Calibration Run} in blue, a \textit{Physics Run} in red and their difference in green, relating to MSD Sensor 4.}
    \label{fig:cum_diff}
\end{figure}

\begin{figure}[htbp] 
    \centering
    \begin{minipage}[b]{0.45\textwidth}
        \centering
        \includegraphics[width=\textwidth]{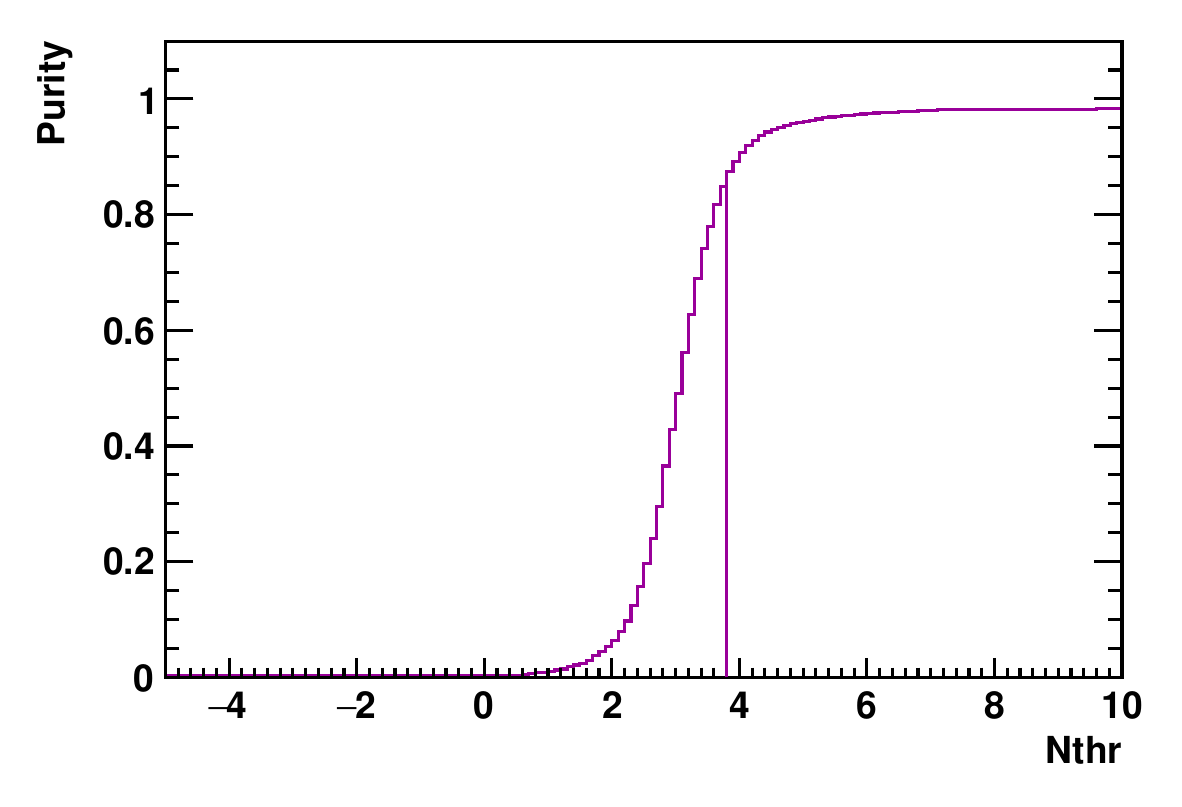}
        \label{fig:subfig_d}
    \end{minipage}
    \hfill
    \begin{minipage}[b]{0.45\textwidth}
        \centering
        \includegraphics[width=\textwidth]{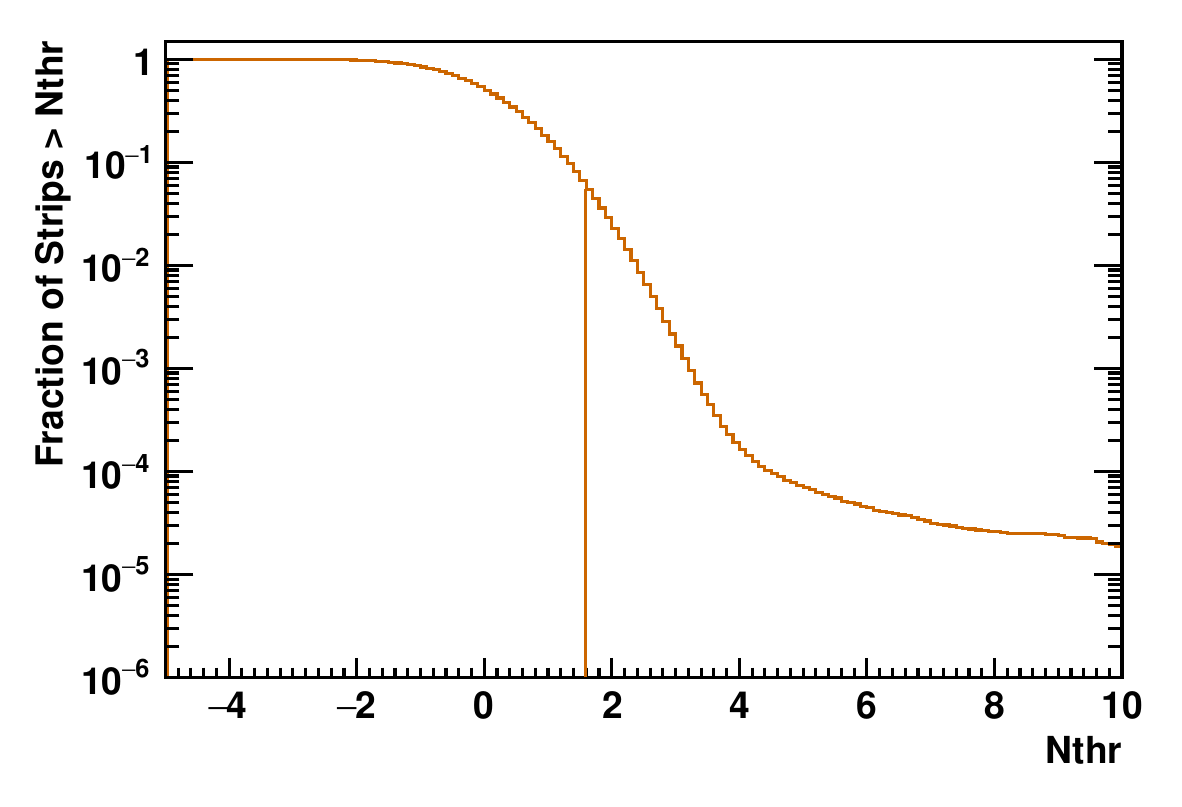}
        \label{fig:subfig_c}
    \end{minipage}
    \caption{ Plots relating to MSD Sensor 4. \textbf{(Left)} The \textit{Purity} curve as a function of the threshold Nthr. For 85$\%$ \textit{Purity}, \textit{Seed Threshold} = 3.9 has been obtained. \textbf{(Right)} The \textit{Fraction of strips > Nthr} curve as a function of the threshold Nthr. For 5$\%$ of \textit{Fake Fired Strips}, \textit{Fired Threshold = } 1.8 has been obtained. }
    \label{fig:soglie}
\end{figure}

After producing the cumulative distributions for both \textit{Calibration} and \textit{Physics Runs}, referred to as CAL and PHY respectively, a third cumulative distribution, named DIFF = PHY - CAL, is obtained by computing the difference bin-by-bin between the physics and calibration histograms and is represented by the green curve in Figure \ref{fig:cum_diff}. The behaviour of the DIFF variable is equal to 0 until the Nthr value increases at the point where the number of relative maxima due to single strip noise fluctuations decrease faster than the number of relative maxima due to particle interactions. This Nthr value is slightly greater than 0, and DIFF increases monotonically until a maximum value is reached. From that point on, DIFF becomes monotonically decrescent due to the reduced capacity of selecting relative maxima, both due to noise fluctuation and true particle interactions. It should be remarked that the analog signal of each strip is only used to define the relative maxima and not to discriminate among true particle interactions and relative maxima due to noise.
\\
At this point, the ad hoc parameter \textit{Purity} is introduced as a function of the threshold Nthr: it represents the fraction of events due to true particle interactions and it is described by the following expression:
\begin{equation}
\text{Pur}(x) = \frac{\text{DIFF}(x)}{\text{PHY}(x)} 
\label{pur_eff}
\end{equation}where \( x \) denotes the threshold value Nthr.
By observing the behavior of the \textit{Purity} as a function of Nthr, as shown in Figure~\ref{fig:soglie} (left), it is possible to perform an initial assignment of the \textit{Seed Threshold} achieving a first estimate of the optimal working point for cluster identification. At this stage of the study, it turned out to be reasonable to require 85$\%$ \textit{Purity}, which, for MSD sensor 4 corresponds to \textit{Seed Threshold} = 3.9.\\\\
Regarding the assignment of the \textit{Fired Threshold} instead, the fraction of potential \textit{Fake Fired Strips} has to be evaluated. To assert this the processed signals of all the strips of a sensor related to a \textit{Calibration Run}, not only of the \textit{Potential Seed Strips}, have to be considered in order to create a cumulative histogram of the \textit{Fraction of strips} having $\text{Signal}_i/\text{SSN}_i$ exceeding a threshold Nthr, as shown in Figure~\ref{fig:soglie} (right). A 5$\%$ of \textit{Fake Fired Strips} is chosen for this study, corresponding to \textit{Fired Threshold} = 1.8.\\\\
 As mentioned above, the case of protons at the highest energy available at the CNAO facility has been presented as they are the particles with the lowest energy deposition among those studied in this work, which makes them the most challenging in terms of cluster identification.
Therefore, the threshold values reported in this paper can be considered a safe and conservative choice when considering particles with greater energy loss.\\
In Table~\ref{tab:thresholds}, the estimated values of the \textit{Seed} and \textit{Fired Thresholds} for all the six sensors of the MSD detector are reported.
\begin{table}[htbp]
    \centering
    \begin{tabular}{ccc}
    \hline
    \textbf{Sensor} & \textbf{Seed Threshold} & \textbf{Fired Threshold} \\
    \hline
    Sensor 0 & 4.7 & 1.9 \\
    \hline
    Sensor 1 & 4.0 & 1.6 \\
    \hline
    Sensor 2 & 3.5 & 1.6 \\
    \hline
    Sensor 3 & 3.6 & 1.7 \\
    \hline
    Sensor 4 & 3.9 & 1.8 \\
    \hline
    Sensor 5 & 5.9 & 1.7 \\
    \hline
\end{tabular}
    \caption{Estimated \textit{Seed} and \textit{Fired Thresholds} for the MSD sensors.} 
    \label{tab:thresholds}
\end{table}
Based on these reference values, a threshold scan can be carried out in their vicinity to evaluate the single-ion detection efficiency of each individual MSD sensor and to orient the clustering study based on tracking information.
\section{Conclusion}
In this work, a new method for setting clustering thresholds in the Microstrip Silicon Detector (MSD) has been introduced, allowing for an estimation of the values independent from the other detectors of the FOOT experiment and providing a systematic framework for threshold optimization. The strategy is based on the comparison of \textit{Potential Seed Strips} of a cluster during data acquisitions both in absence and in presence of particles, and on the evaluation of the number of potential \textit{Fake Fired Strips} in absence of particles.
The analysis has been performed using protons at 230 MeV impinging on the FOOT magnetic spectrometer setup with no target at the CNAO facility, as they are the most challenging particles for clustering identification, due to their minimal ionization. By requiring 85$\%$ \textit{Purity} and 5$\%$ of \textit{Fake Fired Strips}, the \textit{Seed} and \textit{Fired Thresholds} values are set, around which a systematic threshold scan will be performed to evaluate the MSD sensor efficiency exploiting tracking algorithms. This method can be applied to handle data acquisitions involving different Z ions, dealing with the several FOOT experiment data taking campaigns at different facilities.\\The present study lays the groundwork for further improvements in the tracking performance of the MSD, contributing to the overall precision of the FOOT experiment in measuring nuclear fragmentation cross sections.


\begin{thebibliography}{99}

\bibitem{2014CMS_Tracker}
The CMS Collaboration,
\textit{Description and performance of track and primary-vertex reconstruction with the CMS tracker},
Journal of Instrumentation, vol. 9, 2014.
doi: \href{https://doi.org/10.1088/1748-0221/9/10/P10009}{10.1088/1748-0221/9/10/P10009}.

\bibitem{2002AMS_Tracker}
W.J. Burger,
\textit{The AMS silicon tracker},
Nuclear Physics B - Proceedings Supplements, vol. 113, pp. 139--146, 2002.
doi: \href{https://doi.org/10.1016/S0920-5632(02)01833-9}{10.1016/S0920-5632(02)01833-9}.

\bibitem{FOOT_2021}
G. Battistoni, M. Toppi, V. Patera, The FOOT Collaboration,
\textit{Measuring the Impact of Nuclear Interaction in Particle Therapy and in Radio Protection in Space: the FOOT Experiment},
Frontiers in Physics, vol. 8, 2021.
doi: \href{https://doi.org/10.3389/fphy.2020.568242}{10.3389/fphy.2020.568242}.

\bibitem{FOOT-2017}
R. Silvestri, G. Battistoni, S. Bini, et al.,
\textit{The FOOT experiment: Fragmentation of Oxygen ions at intermediate energy},
Nuclear Instruments and Methods in Physics Research Section A, vol. 871, pp. 77--84, 2017.
doi: \href{https://doi.org/10.1016/j.nima.2017.07.014}{10.1016/j.nima.2017.07.014}.

\bibitem{Silvestre_2022}
G. Silvestre, F. Peverini, L. Servoli, The FOOT Collaboration,
\textit{Characterization of 150 $\mu$m thick silicon microstrip prototype for the FOOT experiment},
Journal of Instrumentation, vol. 17, article P12012, 2022.
doi: \href{https://doi.org/10.1088/1748-0221/17/12/P12012}{10.1088/1748-0221/17/12/P12012}.

\bibitem{Kanxheri_2022}
K. Kanxheri, M. Barbanera, G. Ambrosi, G. Silvestre, L. Servoli, The FOOT Collaboration,
\textit{The Microstrip Silicon Detector (MSD) data acquisition system architecture for the FOOT experiment},
Journal of Instrumentation, vol. 17, article C03035, 2022.
doi: \href{https://doi.org/10.1088/1748-0221/17/03/C03035}{10.1088/1748-0221/17/03/C03035}.

\bibitem{Turchetta_1993}
R. Turchetta,
\textit{Spatial resolution of silicon microstrip detectors},
Nucl. Instr. and Meth. A, vol. 335, pp. 44--58, 1993.
doi: \href{https://doi.org/10.1016/0168-9002(93)90255-G}{10.1016/0168-9002(93)90255-G}.

\bibitem{HERAb-2001}
I. Abt et al.,
\textit{Cluster shapes and cluster sizes in the HERA-B silicon vertex detector},
Nuclear Instruments and Methods in Physics Research Section A, vol. 469, pp. 147--158, 2001.
doi: \href{https://doi.org/10.1016/S0168-9002(01)00775-6}{10.1016/S0168-9002(01)00775-6}.

\bibitem{Servoli_2010}
L. Servoli, D. Biagetti, D. Passeri, E. Spanti Gattuso,
\textit{Characterization of standard CMOS pixel imagers as ionizing radiation detectors},
Journal of Instrumentation, vol. 5, article P07003, 2010.
doi: \href{https://doi.org/10.1088/1748-0221/5/07/P07003}{10.1088/1748-0221/5/07/P07003}.

\bibitem{Servoli_2011}
L. Servoli, D. Biagetti, S. Meroli, P. Placidi, D. Passeri, P. Tucceri,
\textit{Use of a standard CMOS imager as position detector for charged particles},
Nuclear Physics B - Proceedings Supplements, vol. 215, pp. 228--231, 2011.
doi: \href{https://doi.org/10.1016/j.nuclphysbps.2011.04.016}{10.1016/j.nuclphysbps.2011.04.016}.


\end{thebibliography}


\end{document}